\documentstyle[amssymb,fleqn,epsf,amsbsy]{article}

\newcommand{\be}{\begin{equation}}
\newcommand{\ee}{\end{equation}}
\newcommand{\ba}{\begin{array}}
\newcommand{\ea}{\end{array}}
\newcommand{\bea}{\begin{eqnarray}}
\newcommand{\eea}{\end{eqnarray}}
\newcommand{\bean}{\begin{eqnarray*}}
\newcommand{\eean}{\end{eqnarray*}}

\newcommand{\lp}{\left(}
\newcommand{\rp}{\right)}
\newcommand{\ls}{\left[}
\newcommand{\rs}{\right]}
\newcommand{\lc}{\left\{}
\newcommand{\rc}{\right\}}

\newcommand{\la}{\langle}

\newcommand{\ra}{\rangle}

\renewcommand{\d}{\mbox{\rm d}}
\newcommand{\e}{\mbox{\rm e}}
\newcommand{\p}{\partial}
\newcommand{\intl}{\int\limits}
\newcommand{\suml}{\sum\limits}
\newcommand{\im}{{\rm i}}
\newcommand{\veps}{\varepsilon}
\newcommand{\vphi}{\varphi}
\newcommand{\vrho}{\varrho}

\newcommand{\mGamma}{{\mit\Gamma}}
\newcommand{\mDelta}{{\mit\Delta}}
\newcommand{\mTheta}{{\mit\Theta}}
\newcommand{\mPhi}{{\mit\Phi}}
\newcommand{\mOmega}{{\mit\Omega}}

\newcommand{\const}{\mathop{{\rm const}}\nolimits}
\newcommand{\ddt}{\frac{\partial}{\partial t}}
\newcommand{\dd}[1]{\frac{\partial}{\partial #1}}
\newcommand{\refp}[1]{(\ref{#1})}
\newcommand{\ds}{\displaystyle}

\newcommand{\bmv}[1]{\mbox{\boldmath $#1$}}
\newcommand{\ohs}[2]{{}^{#1}\mOmega_{\rm hs}^{#2}}
\newcommand{\ol}[2]{{}^{#1}\mOmega_{\rm l}^{#2}}
\newcommand{\tensor}[1]{\stackrel{\leftrightarrow}{#1}{}}

\renewcommand{\thesection}{\arabic{section}.}

\title{\Large\bf Normal solution and transport coefficients\\ to the 
Enskog-Landau kinetic equation\\ for a two-component system of charged hard 
spheres.\\ The Chapman-Enskog method}
\author{\small A.E.Kobryn, I.P.Omelyan, M.V.Tokarchuk\\ 
\small Institute for Condensed Matter Physics, Ukrainian National Academy of 
Sciences,\\ 
\small 1~Svientsitskii~St., UA--290011 Lviv--11, Ukraine}
\date{\normalsize\today}

\begin{document}
\maketitle

\begin{abstract}
An Enskog-Landau kinetic equation for a many-component system of 
charged hard spheres is proposed. It has been obtained from the Liouville 
equation with modified boundary conditions by the method of nonequilibrium 
statistical operator. On the basis of this equation the normal solutions 
and transport coefficients such as bulk $\kappa$ and shear $\eta$ 
viscosities, thermal conductivity $\lambda$, mutual diffusion 
$D^{\alpha\beta}$ and thermal diffusion $D_{\rm T}^\alpha$ have been 
obtained for a binary mixture in the first approximation using the 
Chapman-Enskog method. Numerical calculations of $\kappa$, $\eta$, 
$D^{\alpha\beta}$ and $D_{\rm T}^\alpha$ for mixtures Ar-Kr, Ar-Xe, Kr-Xe 
with different concentrations of compounds have been evaluated for the cases 
of absence and presence of long-range Coulomb interactions. The results 
are compared with those obtained from other theories and experiment.
\end{abstract}

\section{Introduction}
\setcounter{equation}{0}

The Enskog-Landau kinetic equation for an one-component system of charged 
hard spheres \cite{c1} is comparatively new. Being obtained from the first 
principles of statistical mechanics it describes well transport processes in 
moderately dense systems. Its collision integral does not have any 
singularities on short-range distances. Numerical calculations of shear 
viscosity $\eta$ and thermal conductivity $\lambda$ on the basis of normal 
solutions to this equation using the Chapman-Enskog method are in a good 
agreement between theory and experiment \cite{c2}. In next papers 
\cite{c3p,c3}, nonstationary solutions and transport coefficients have been 
obtained using the boundary conditions method. In the limiting case of 
stationary processes, the results of \cite{c2} and \cite{c3} coincide 
between themselves. Despite this fact, the one-component system remains 
always by a model system \cite{c4a,c4b,c4c}.

Traditional interest in many-component systems, including two-component 
mixtures of dense gases, liquids and plasmas, has been intensified recently 
by the considerable advancement in computer equipment. This allows to 
supplement theory by results of mathematical simulations rather easily.

The transport coefficients of nonequilibrium systems have been investigated 
using different methods, namely: Bhatnagar-Gross-Kruk theory \cite{c5}, 
Green-Kubo \cite{c6a} and Frost \cite{c7} methods, at unified 
description of kinetics and equations of nonlinear hydrodynamics \cite{c8}, 
in one-liquid approximation \cite{c9}. Mathematical construction of 
extended thermodynamics of dense gases and liquids (including hard sphere 
mixtures within RET theory \cite{c1}) has been carried out \cite{c11}, 
thermodynamical compatibility of Enskog-type kinetic equations for 
$M$-component systems and Enskog-like kinetic equations for reacting 
mixtures has been proved \cite{c12}, results of macroscopic and 
microscopic theories have been compared \cite{c14}. At the same time, 
when methods for calculating transport coefficients were improved, 
numerical simulations such as Monte Carlo (MC) and molecular dynamics (MD) 
were optimized \cite{c16,c17,c18} also. In some cases, purely theoretical 
problems lead to some other ones: the Gross-Jackson model \cite{c19}, 
inverse problem of thermal conductivity \cite{c20}, effective transport 
coefficients \cite{c21,c22,c23}. In such a way the efforts to expand the 
existing theory \cite{c1,c2} on many-component systems looks quite naturally. 
Moreover, a sequential theory for charged particles does not exist yet. In 
papers \cite{c5,c6a,c7,c8,c9,c11,c12,c14,c16,c17,c18,c19,%
c20,c21,c22,c23}, transport coefficients (such as viscosity $\eta$, 
conductivity  $\sigma$, thermal conductivity $\lambda$, mutual diffusion 
$D^{\alpha\beta}$ and thermal diffusion $D_{\rm T}^\alpha$) have been 
calculated in one or other way. It is pointed everywhere, that the best 
coincidence of theoretical and experimental data (as well as MC or MD) is 
obtained for neutral systems at comparatively not high densities. As was 
noted in \cite{c17}, the coincidence at high densities is bad (theory does 
not work). It is summarized in \cite{c18} on the basis of MC simulations, 
that $\eta$ coincides with experimental value better than $\lambda$ in major 
cases. Generally, a worse agreement of theory and experiment for $\lambda$ 
is known for a long time. First of all, it is connected with a strong 
growing of interparticle correlations as density decreases. These 
correlations should be taken into account more precisely \cite{c24}. In its 
turn, this leads to changes in local conservation laws, which are used 
at finding for solutions to kinetic equations \cite{c3}.

In the present paper, the Enskog-Landau kinetic equation for a 
multicomponent system of charged hard spheres, its normal solution by means 
of the Chapman-Enskog method, and the calculation of transport coefficients 
for binary mixtures are proposed. The theoretical part of this work was 
published for the first time in papers \cite{c25,c26}. Now it is upgraded 
by some results of numerical calculations for transport coefficients as well 
as their comparison with some experiments and MD data. A special attention 
is for $\mOmega$-integrals calculation for long-range type of interaction 
(Coulomb interaction in our case) and for finding of static screening 
radius. Expressions for spatially dependent distribution functions in 
relations for transport coefficients are discussed concisely as well.

\section{Initial relations}
\setcounter{equation}{0}

Let us consider a nonequilibrium system of $N$ classical particles, which 
consists of $M$ kinds of charged hard spheres. Each kind has $N_\alpha$ 
particles, where $\alpha$ is varied between 1 and $M$, with 
$\sum_{\alpha=1}^MN_\alpha=N$. The masses of particles are equal to 
$m_\alpha$, altogether they occupy a volume $V$. It is agreed that the 
thermodynamic limit transition $N\to\infty$, $V\to\infty$, $N/V=\const$ 
takes place. The charges of particles are equal to $Z_\alpha e$, where $e$ 
is the electron charge, $Z_\alpha\in{\Bbb Z}$. The condition of 
electroneutrality is applied to the system as a whole. It means that in the 
case of differently charged particles 
$\sum_{\alpha=1}^MN_\alpha Z_\alpha=0$, while in the opposite cases, where 
only positively charged particles are considered (mixtures of ionized gases, 
for example), a neutralizing continuum is introduced.

The Hamiltonian of such a system can be presented as follows:
\[
H=\sum_{\alpha=1}^M\sum_{j_\alpha=1}^{N_\alpha}
	\frac{p_{j_\alpha}}{2m_\alpha}+\frac 12
	\mathop{\sum_{\alpha=1}^M\sum_{\beta=1}^M
	\sum_{j_\alpha=1}^{N_{\mathstrut\alpha}}
	\sum_{k_\beta=1}^{N_{\mathstrut\beta}}}
	\limits_{j_\alpha\ne k_\beta\;\mbox{\scriptsize at}\;\alpha=\beta}
	\mPhi(|\bmv{r}_{j_\alpha}-\bmv{r}_{k_\beta}|),
\]
where
\[
\mPhi(|\bmv{r}_{j_\alpha}-\bmv{r}_{k_\beta}|)=
	\mPhi(|\bmv{r}_{j_\alpha k_\beta}|)=
	\mPhi^{\rm hs}(|\bmv{r}_{j_\alpha k_\beta}|)+
	\mPhi^{\rm l}(|\bmv{r}_{j_\alpha k_\beta}|)
\]
is the total interparticle interaction potential, which consists of the 
short-range (hard spheres) $\mPhi^{\rm hs}$ and the long-range (Coulomb one 
in our case) $\mPhi^{\rm l}$ interactions:
\[
\mPhi^{\rm hs}(|\bmv{r}_{j_\alpha k_\beta}|)=\lim_{c\to\infty}
	\mPhi^c(|\bmv{r}_{j_\alpha k_\beta}|)=
\lc
\ba{ll}
 c,&|\bmv{r}_{j_\alpha k_\beta}|<\sigma_{\alpha\beta},\\
 0,&|\bmv{r}_{j_\alpha k_\beta}|>\sigma_{\alpha\beta},\\
\ea\right.
\]
\[
\mPhi^{\rm l}(|\bmv{r}_{j_\alpha k_\beta}|)=\lc
\ba{ll}
 0,&|\bmv{r}_{j_\alpha k_\beta}|<\sigma_{\alpha\beta},\\
\ds\frac{Z_\alpha Z_\beta e^2}{|\bmv{r}_{j_\alpha k_\beta}|},
&|\bmv{r}_{j_\alpha k_\beta}|>\sigma_{\alpha\beta}.\\
\ea\right.
\]

A nonequilibrium state of this system is described by the nonequilibrium 
$N$-particle distribution function $\varrho\lp x^N;t\rp$. Let us consider 
now much precisely the case of a two-component system, when $M=2$. Then the 
nonequilibrium $N$-particle distribution function, 
\bean
\vrho\lp x^N;t\rp&=&\vrho\lp(x^a)^{N_a},(x^b)^{N_b};t\rp\\
&=&\vrho\lp x_1^a,x_2^a,\ldots,x_{N_a}^a,
	x_1^b,x_2^b,\ldots,x_{N_b}^b;t\rp,
\eean
satisfies the Liouville equation and normalization condition \cite{c1,c2}. 
In the two-component case the $N$-particle Liouville operator has a 
different structure, which takes into account a many-component nature of the 
system:
\[
L_N=\sum_{\alpha=1}^M\sum_{j_\alpha=1}^{N_\alpha}L(j_\alpha)+\frac 12
	\mathop{\sum_{\alpha=1}^M\sum_{\beta=1}^M
	\sum_{j_\alpha=1}^{N_{\mathstrut\alpha}}
	\sum_{k_\beta=1}^{N_{\mathstrut\beta}}}
	\limits_{j_\alpha\ne k_\beta\;\mbox{\scriptsize при}\;\alpha=\beta}
	L(j_\alpha,k_\beta),
\]
\[
L(j_\alpha)=-\im\frac{\bmv{p}_{j_\alpha}}{m_\alpha}
	\dd{\bmv{r}_{j_\alpha}},
\]
\[
L(j_\alpha,k_\beta)=
	\im\frac{\p\mPhi(|\bmv{r}_{j_\alpha}-\bmv{r}_{k_\beta}|)}
	{\p\bmv{r}_{j_\alpha}}
	\lp\dd{\bmv{p}_{j_\alpha}}-\dd{\bmv{p}_{k_\beta}}\rp.
\]

\section{Kinetic equation}
\setcounter{equation}{0}

The kinetic equation for $\alpha$-kind one-particle distribution function 
$f_1(x_1^\alpha;t)$ can be obtained similarly to an one-component system, 
but with some distinguishes, which are due to a multicomponent nature of the 
system under consideration. In such a way, in the ``pair'' collisions 
approximation this equation reads:
\bea
\lefteqn{\ds\lp\ddt+\im L(1_\alpha)\rp f_1(x_1^\alpha;t)=
	-\sum_{\beta=1}^M\int\d x_2^\beta\;\im L(1_\alpha,2_\beta)
	\times{}}\label{e3.2.1}\\
&&\lim_{\tau\to-\infty}\e^{\ds\im L_2\tau}
	g_2(\bmv{r}_1^\alpha,\bmv{r}_2^\beta;t+\tau)
	f_1(x_1^\alpha;t+\tau)f_1(x_2^\beta;t+\tau),\nonumber
\eea
where
\[
g_2(\bmv{r}_1^\alpha,\bmv{r}_2^\beta;t)=
	\big[n(\bmv{r}_1^\alpha;t)n(\bmv{r}_2^\beta;t)\big]^{-1}
	f_2(\bmv{r}_1^\alpha,\bmv{r}_2^\beta;t),
\]
\[
f_2(\bmv{r}_1^\alpha,\bmv{r}_2^\beta;t)=
	\int\d\mGamma'_N\;n_2(\bmv{r}_1^\alpha,\bmv{r}_2^\beta(x')^N;t)
	\vrho_q\lp(x')^N;t\rp.
\]
Equation (\ref{e3.2.1}) is formally distinguished on analogous one for an 
one-component system by the presence of kind summations on its right-hand 
part in the collision integral. The Enskog-Landau kinetic equation for a 
two-component system of charged hard spheres has been obtained for the first 
time in \cite{c25}:
\bea
\lefteqn{\ds\ls\ddt+\im L(1_\alpha)\rs f_1(x_1^\alpha;t)={}}
	\label{e3.2.2}\\
&&I_{\rm E}^{(0)}(x_1^\alpha;t)+I_{\rm E}^{(1)}(x_1^\alpha;t)+
	I_{\rm MF}(x_1^\alpha;t)+I_{\rm L}(x_1^\alpha;t),\nonumber
\eea
where terms in the right-hand part are collision integrals. Explicit 
expressions for the collision integrals depend on a specific interparticle 
potential of interaction. The first and the second terms are collision 
integrals of the revised Enskog theory (RET) \cite{c1,c2}:
\bea
\lefteqn{\ds I_{\rm E}^{(0)}(x_1^\alpha;t)=\sum_{\beta=1}^M
	\int\d\bmv{v}_2^\beta\;
	\d\veps\;\d b\;b\;g^{\alpha\beta}\;
	g_2^{\alpha\beta}(\sigma_{\alpha\beta}|n,\beta)\times{}}
	\label{e3.2.3}\\
&&\Big(f_1(\bmv{r}_1^\alpha,\bmv{v}_1^{\alpha\prime};t)
	f_1(\bmv{r}_2^\beta,\bmv{v}_2^{\beta\prime};t)-
	f_1(\bmv{r}_1^\alpha,\bmv{v}_1^{\alpha};t)
	f_1(\bmv{r}_2^\beta,\bmv{v}_2^{\beta};t)\Big),\nonumber
\eea
\bea
\lefteqn{\ds I_{\rm E}^{(1)}(x_1^\alpha;t)=
	\sum_{\beta=1}^M\sigma_{\alpha\beta}^3
	\int\d\hat{\bmv{r}}_{12}^{\alpha\beta}\;
	\d\bmv{v}_{2}^{\beta}\;
	\mTheta(\hat{\bmv{r}}_{12}^{\alpha\beta}\cdot\bmv{g}^{\alpha\beta})
	(\hat{\bmv{r}}_{12}^{\alpha\beta}\cdot\bmv{g}^{\alpha\beta})\times{}}
	\nonumber\\
&&\bigg[\hat{\bmv{r}}_{12}^{\alpha\beta}
	g_2^{\alpha\beta}(\bmv{r}_{12}^{\alpha\beta}|n,\beta)\times{}
	\label{e3.2.4}\\
&&\Big[f_1(\bmv{r}_1^\alpha,\bmv{v}_1^{\alpha\prime};t)\boldsymbol\nabla
	f_1(\bmv{r}_2^\beta,\bmv{v}_2^{\beta\prime};t)-
	f_1(\bmv{r}_1^\alpha,\bmv{v}_1^{\alpha};t)\boldsymbol\nabla
	f_1(\bmv{r}_2^\beta,\bmv{v}_2^{\beta};t)\Big]+{}\nonumber\\
&&\frac 12\ls\hat{\bmv{r}}_{12}^{\alpha\beta}\cdot\boldsymbol\nabla
	g_2^{\alpha\beta}(\bmv{r}_{12}^{\alpha\beta}|n,\beta)\rs
	\times{}\nonumber\\
&&\Big[f_1(\bmv{r}_1^\alpha,\bmv{v}_1^{\alpha\prime};t)
	f_1(\bmv{r}_2^\beta,\bmv{v}_2^{\beta\prime};t)+
	f_1(\bmv{r}_1^\alpha,\bmv{v}_1^{\alpha};t)
	f_1(\bmv{r}_2^\beta,\bmv{v}_2^{\beta};t)\Big]\bigg].\nonumber
\eea
The next term is caused by the influence of long-range interactions in 
the mean field approximation (KMFT) \cite{c1,c2}:
\bea
\lefteqn{\ds I_{\rm MF}(x_1^\alpha;t)={}}\label{e3.2.5}\\
\lefteqn{\ds\frac{1}{m_\alpha}
	\sum_{\beta=1}^M\int\d\bmv{r}_2^\beta\;
	\dd{\bmv{r}_{1}^{\alpha}}
	\mPhi^{\rm l}(|\bmv{r}_{12}^{\alpha\beta}|)
	g_2^{\alpha\beta}(\bmv{r}_{1}^{\alpha},\bmv{r}_{2}^{\beta};t)
	\dd{\bmv{v}_{1}^{\alpha}}f_1(x_1^\alpha;t)
	n_\beta(\bmv{r}_2^\beta;t),}\nonumber
\eea
the last term is a Landau-like collision integral \cite{c1,c2}:
\bea
\lefteqn{\ds I_{\rm L}(x_1^\alpha;t)=\sum_{\beta=1}^M\int\d\bmv{v}_2^\beta\;
	\d\veps\;\d b\;b\;g^{\alpha\beta}\;\times{}}\label{e3.2.6}\\
&&\ds\Big(f_1(\bmv{r}_1^\alpha,\bmv{v}_1^{\alpha *};t)
	f_1(\bmv{r}_2^\beta,\bmv{v}_2^{\beta *};t)-
	f_1(\bmv{r}_1^\alpha,\bmv{v}_1^{\alpha};t)
	f_1(\bmv{r}_2^\beta,\bmv{v}_2^{\beta};t)\Big).\nonumber
\eea
This collision integral is presented here in a reduced Boltzmann-like form. 
It could be obtained going from the cartesian to cylindrical reference 
frame, entering impact parameter $b$, azimuthal angle of scattering 
$\varepsilon$, distance along cylinder centreline $\xi$ and integrating 
with respect to $\xi$ and taking into account the condition 
$g_2(\bmv{r}_1,\bmv{r}_2;t) \to 1$. If one solves this kinetic equation 
using the Chapman-Enskog method, this form for the latest collision integral 
is more convenient \cite{c2}. Other conventional designations in 
(\ref{e3.2.3}) -- (\ref{e3.2.6}) are as follows:
\[
\ba{ll}
\itemsep=-2pt
b&\mbox{impact parameter,}\\
\beta&\mbox{inversed local temperature analogue,}\\
\veps&\mbox{azimuthal angle of scattering,}\\
\bmv{g}^{\alpha\beta}&\mbox{relative velocity of $\alpha$- and 
$\beta$-kind particles,}\\
g_2^{\alpha\beta}&\mbox{two-particle correlation function,}\\
m^*&\mbox{reduced mass,}\\
m_\alpha&\mbox{partial masses of particles,}\\
n&\mbox{total density of particles number,}\\
n_\alpha&\mbox{partial densities of particles numbers,}\\
\hat{\bmv{r}}_{12}^{\alpha\beta}&\mbox{unit vector along 
	$\bmv{r}_{12}^{\alpha\beta}$ direction,}\\
\mTheta(x)&\mbox{unit step function.}\\
\ea
\]
An additional point for emphasizing is that $\bmv{v}_1^{\alpha\prime}$ and 
$\bmv{v}_2^{\alpha\prime}$ are hard spheres velocities after collision:
\bea
\bmv{v}_1^{\alpha\prime}&=&\bmv{v}_1^{\alpha}+
	\hat{\bmv{r}}_{12}^{\alpha\beta}
	(\hat{\bmv{r}}_{12}^{\alpha\beta}\cdot\bmv{g}^{\alpha\beta}),
	\nonumber\\
\bmv{v}_2^{\beta\prime}&=&\bmv{v}_2^{\beta}-
	\hat{\bmv{r}}_{12}^{\alpha\beta}
	(\hat{\bmv{r}}_{12}^{\alpha\beta}\cdot\bmv{g}^{\alpha\beta}),
	\nonumber
\eea
as well as $\bmv{v}_1^{\alpha *}$ and $\bmv{v}_2^{\alpha *}$ are velocities 
of charged particles after Coulomb scattering:
\bean
\bmv{v}_1^{\alpha *}&=&\bmv{v}_1^{\alpha}+
	\vartriangle\!\!\bmv{v}^{\alpha\beta},\\
\bmv{v}_2^{\beta *}&=&\bmv{v}_1^{\beta}-
	\vartriangle\!\!\bmv{v}^{\alpha\beta},
\eean
\be
\vartriangle\!\!\bmv{v}^{\alpha\beta}=-\frac{1}{m^*}
	\int\d\xi\;\dd{\bmv{r}_{12}^{\alpha\beta}}
	\mPhi^{\rm l}(|\bmv{r}_{12}^{\alpha\beta}|)
	\frac{1}{g^{\alpha\beta}}
	\Bigg|_{\ds r_{12}^{\alpha\beta}=\sqrt{b^2+\xi^2}}.
	\label{e3.7}
\ee

\section{Normal solutions by means of the Chapman-Enskog method in the first 
approximation}
\setcounter{equation}{0}

As usually, to a solve kinetic equation by means of the Chapman-Enskog 
method in the $k\,$th approximation we will use conservation laws for a set 
of hydrodynamical variables in the $(k-1)\,$th approximation. The components 
of the additive invariant vector could be chosen similarly to \cite{c2,c3}, 
namely the hydrodynamical mass density, momentum and energy:
\bean
\lefteqn{\ds\rho(\bmv{r}_1;t)}\qquad\qquad\qquad\quad&=
	&\sum_{\alpha=1}^M\int\d\bmv{v}_1^\alpha\;f_1(x_1^\alpha;t)
	m_\alpha,\nonumber\\
\lefteqn{\ds\rho(\bmv{r}_1;t)\bmv{V}(\bmv{r}_1;t)}\qquad\qquad\qquad\quad&=
	&\sum_{\alpha=1}^M\int\d\bmv{v}_1^\alpha\;f_1(x_1^\alpha;t)
	m_\alpha\bmv{v}_1^\alpha,\nonumber\\
\lefteqn{\ds\rho(\bmv{r}_1;t)
	\omega_{\rm k}(\bmv{r}_1;t)}\qquad\qquad\qquad\quad&=
	&\sum_{\alpha=1}^M\int\d\bmv{v}_1^\alpha\;f_1(x_1^\alpha;t)
	\frac{m_\alpha\lp\bmv{c}_1^\alpha\rp^2}{2}.\nonumber\\
\eean
The conservation law for $\rho(\bmv{r}_1;t)$ have the form of a continuity 
equation, whereas the equation of motion and equation of energy balance read 
as
\bea
\rho(\bmv{r}_1;t)\frac{\d\bmv{V}(\bmv{r}_1;t)}{\d t}&=
	&\!\!-\dd{\bmv{r}_1}\tensor{P}'(\bmv{r}_1;t),\label{e3.2.7}\\
\rho(\bmv{r}_1;t)\frac{\d\omega_{\rm k}(\bmv{r}_1;t)}{\d t}&=
	&\!\!-\dd{\bmv{r}_1}\bmv{q}^*(\bmv{r}_1;t)-
	\tensor{P}^*(\bmv{r}_1;t):\dd{\bmv{r}_1}\bmv{V}(\bmv{r}_1;t)+{}
	\nonumber\\
\lefteqn{\ds\sum_{\alpha,\beta}^M\bmv{V}_{\rm d}^\alpha(\bmv{r}_1;t)
	\int\d\bmv{r}_{12}^{\alpha\beta}\;
	\dd{r_{12}^{\alpha\beta}}
	\mPhi^{\rm l}\lp|\bmv{r}_{12}^{\alpha\beta}|\rp
	n_\alpha(\bmv{r}_1^\alpha;t)
	n_\beta(\bmv{r}_1^\alpha+\bmv{r}_{12}^{\alpha\beta};t),}
	\qquad\qquad\qquad\;\nonumber
\eea
where
\bean
\lefteqn{\ds\tensor{P}'(\bmv{r}_1;t)}\qquad\qquad&=
	&\tensor{P}^{\rm k}(\bmv{r}_1;t)+\tensor{P}^{\rm hs}(\bmv{r}_1;t)+
	\tensor{P}^{\rm mf}(\bmv{r}_1;t),\nonumber\\
\lefteqn{\ds\tensor{P}^*(\bmv{r}_1;t)}\qquad\qquad&=
	&\tensor{P}^{\rm k}(\bmv{r}_1;t)+\tensor{P}^{\rm hs}(\bmv{r}_1;t),\\
\lefteqn{\ds\bmv{q}^*(\bmv{r}_1;t)}\qquad\qquad&=
	&\bmv{q}^{\rm k}(\bmv{r}_1;t)+\bmv{q}^{\rm hs}(\bmv{r}_1;t);\\
\eean
$\bmv{V}_{\rm d}^\alpha(\bmv{r}_1;t)$ is the diffusion velocity of 
$\alpha$-kind particles:
\[
\bmv{V}_{\rm d}^\alpha(\bmv{r}_1;t)=\la\bmv{c}^\alpha(\bmv{r}_1;t)\ra,
	\qquad\bmv{c}^\alpha(\bmv{r}_1;t)=\bmv{v}_1^\alpha-
	\bmv{V}(\bmv{r}_1;t),
\]
here $\la\ldots\ra$ denotes an average value over the one-particle 
distribution function $f_1(x_1^\alpha;t)$. In the zeroth approximation, when 
the function $f_1(x_1^\alpha;t)$ is taken as a local-equilibrium Maxwell 
distribution function ($k$ is the Boltzmann constant),
\[
f_1^{(0)}(x_1^\alpha;t)=n_\alpha(\bmv{r}_1^\alpha;t)
	\lp\frac{m_\alpha}{2\pi kT(\bmv{r}_1^\alpha;t)}\rp^{3/2}
	\exp\lc-\frac{m_\alpha\big(c_1^\alpha(\bmv{r}_1^\alpha;t)\big)^2}
	{2kT(\bmv{r}_1^\alpha;t)}\rc,
\]
one obtains for $\tensor{P}'(\bmv{r}_1;t)$, $\bmv{q}^*(\bmv{r}_1;t)$,
$\bmv{V}_{\rm d}^\alpha(\bmv{r}_1;t)$ the following expressions:
\be
\tensor{P}^{\rm k}=\tensor{I}P^{\rm k},
	\quad P^{\rm k}=\sum_{\alpha=1}^Mn_\alpha kT,\label{e3.2.8}
\ee
\[
\tensor{P}^{\rm hs}=\tensor{I}P^{\rm hs},\quad P^{\rm hs}=\frac 23
	\pi kT\sum_{\alpha,\beta}^M n_\alpha n_\beta
	\sigma^3_{\alpha\beta}g_2^{\alpha\beta}\left(\sigma|n,\beta\right),
\]
\[
\tensor{P}^{\rm mf}=\tensor{I}P^{\rm mf},\;P^{\rm mf}=-\frac 23\pi
	\sum_{\alpha,\beta}^M n_\alpha n_\beta
	\int\limits_{\sigma_{\alpha\beta}}^\infty
	\d x\; x^3 \lp\mPhi^{\rm l}_{\alpha\beta}(x)\rp'
	g_2^{\alpha\beta}(x),
\]
\[
\tensor{P}^{\rm l}=0,\quad\bmv{q}^{\rm k}=\bmv{q}^{\rm hs}=\bmv{q}^{\rm mf}=
	\bmv{q}^{\rm l}=0,\quad\bmv{V}_{\rm d}^\alpha(\bmv{r}_1;t)=0.
\]
Using the mass conservation law and equation \refp{e3.2.7} in the zeroth 
approximation and taking into consideration an explicit form for 
$\tensor{P}'(\bmv{r}_1;t)$, $\tensor{P}^*(\bmv{r}_1;t)$, 
$\bmv{q}^*(\bmv{r}_1;t)$ \refp{e3.2.8}, we will find for the solution to 
equation \refp{e3.2.2} in the first approximation as the combination:
\be
f_1(x_1^\alpha;t)=f_1^{(0)}(x_1^\alpha;t)
	\big(1+\vphi(x_1^\alpha;t)\big),\label{e3.2.9}
\ee
where the correction $\vphi(x_1^\alpha;t)$ is expressed through 
the Sonine-Laguerre polynomials:
\bean
\lefteqn{\ds\vphi(x_1^\alpha;t)}\qquad\qquad\;
&=&\sqrt{\frac{m_\alpha}{2kT}}
	E^\alpha\lp\frac{m_\alpha\lp{c}_1^\alpha\rp^2}{2kT}\rp
	n\lp\bmv{c}_1^\alpha,\bmv{d}^\alpha\rp-{}\nonumber\\
&&\sqrt{\frac{m_\alpha}{2kT}}
	A^\alpha\lp\frac{m_\alpha\lp{c}_1^\alpha\rp^2}{2kT}\rp
	\lp\bmv{c}_1^\alpha,\boldsymbol\nabla\rp\ln T-{}\nonumber\\
&&\frac{m_\alpha}{2kT}
	B^\alpha\lp\frac{m_\alpha\lp{c}_1^\alpha\rp^2}{2kT}\rp
	\Big(\bmv{c}_1^\alpha\bmv{c}_1^\alpha-
	\frac 13(c_1^\alpha)^2\tensor{I}\Big):
	{\boldsymbol\nabla}{\boldsymbol V}.\nonumber
\eean
Here
\bea
\lefteqn{\ds E^\alpha(x)=\sum_{n=0}^{\infty}
	E_n^\alpha L_n^{3/2}(x),}\nonumber\\
\lefteqn{\ds A^\alpha(x)=\sum_{n=0}^{\infty}
	A_n^\alpha L_n^{3/2}(x),\qquad
	B^\alpha(x)=\sum_{n=0}^{\infty}
	B_n^\alpha L_n^{5/2}(x),}\nonumber\\
\lefteqn{\ds L_n^{r}(x)=\sum_{m=0}^n(-1)^mx^m
	\frac{n!\mGamma(n+r+1)}{m!\mGamma(m+r+1)\mGamma(n-m+1)},}
	\nonumber
\eea
$\bmv{d}^\alpha\equiv\bmv{d}^\alpha(\bmv{r}_1^\alpha;t)$ is the diffusion 
thermodynamic force of $\alpha$-kind of mixture \cite{c27}. Quantities 
$\bmv{d}^\alpha$ satisfy the condition
\[
\sum_{\alpha=1}^M\bmv{d}^\alpha(\bmv{r}_1^\alpha;t)=0.
\]
In the case of an one-component system, diffusion thermodynamic forces are 
absent.

If one puts the solution of \refp{e3.2.9} into equation \refp{e3.2.2}, one 
shall obtain a nonuniform integral equation of Fredgolm-type. Therefore, 
additional conditions for the existence of solutions to this equation can be 
written down at once: this is orthogonality of the right-hand side of 
\refp{e3.2.2} up to solutions of the corresponding uniform equation. The 
last ones are parameters of the abbreviated hydrodynamic description. 
Whereas, additional conditions after mathematical transformations read:
\[
E^b_0=-\frac{n_a}{n_b}\sqrt{\frac{m_a}{m_b}}E^a_0,\qquad
A^b_0=-\frac{n_a}{n_b}\sqrt{\frac{m_a}{m_b}}A^a_0.
\]
This should be taken into consideration any time one looks for expansion 
coefficients $E_n^\alpha$, $A_n^\alpha$, $B_n^\alpha$. In such a way, 
several independent infinite systems of equations appear. It is impossible, 
of course, to find an exact solution to these systems. One should restrict 
by a few first terms in expansion series. As a rule, practically, one 
considers quantities $E_0^\alpha$, $B_0^\alpha$ and $A_0^\alpha$, 
$A_1^\alpha$ only. We should take in mind the fact, that the collision of 
identical particles gives no affect due to mass, momentum and energy 
conservation laws at elastic scattering.

The basic relations to obtain $A_0^\alpha$ and $A_1^\alpha$ coefficients as 
well as their symbolic solution are presented in \ref{aa} The same for 
$B_0^\alpha$ coefficients is done in \ref{ab}

\section{Transport coefficients}
\setcounter{equation}{0}

Having one-particle distribution function \refp{e3.2.9} in the first 
approximation, we are able to write down the expression for mass density 
flow of $\alpha$-component of mixture \cite{c28}:
\[
\bmv{j}^\alpha(\bmv{r}_1^\alpha;t)=-
	\rho^{-1}n^2m_\alpha m_\beta D^{\alpha\beta}
	\bmv{d}^\alpha(\bmv{r}_1^\alpha;t)-
	D_{\rm T}^\alpha\boldsymbol\nabla\ln T(\bmv{r}_1^\alpha;t),
\]
where $D^{\alpha\beta}$ is the mutual diffusion coefficient, and 
$D_{\rm T}^\alpha$ is the thermal diffusion coefficient of $\alpha$-kind of 
mixture. In the zeroth polynomial approximation
\be
D^{\alpha\beta}=-\frac{n_\alpha\rho}{m_\beta n}
	\sqrt{\frac{kT}{2m_\alpha}}E^\alpha_0,\label{e3.2.10}
\ee
\be
D_{\rm T}^\alpha=m_\alpha n_\alpha\sqrt{\frac{kT}{2m_\alpha}}A^\alpha_0.
	\label{e3.2.11}
\ee
Then the expression for thermal diffusion ratio is easy to write as
\[
k_{\rm T}=\frac{\rho}{n^2m_\alpha m_\beta}\frac{D_{\rm T}}{D^{\alpha\beta}}=
	-\frac{1}{n}\frac{A^\alpha_0}{E^\alpha_0}.
\]
Final expressions for coefficients of thermal diffusion one can write taking 
into account the structure of $A^\alpha_0$-coefficients (\ref{aa}). To 
calculate quantities $E^\alpha_0$ it is convenient to consider new 
variables, namely, the variables of inertia centre. By that transformations 
these quantities are calculated exactly. This allows to write the expression 
for mutual diffusion coefficient in an analytical form:
\bean
E^\alpha_0&=&-\frac{3\pi m_\beta}{8\rho n_\alpha}
	\sqrt{\frac{\pi m_\alpha}{m^*}}
	\lp g_2^{\alpha\beta}(\sigma_{\alpha\beta}|n,\beta)\;
	\ohs{\alpha\beta}{(1,1)}+
	\ol{\alpha\beta}{(1,1)}\rp^{-1}\!\!\!,\\
D^{\alpha\beta}&=&\frac{3\pi}{8n}\sqrt{\frac{\pi kT}{2m^*}}
	\lp g_2^{\alpha\beta}(\sigma_{\alpha\beta}|n,\beta)\;
	\ohs{\alpha\beta}{(1,1)}+
	\ol{\alpha\beta}{(1,1)}\rp^{-1}.
\eean
Here $\ohs{\alpha\beta}{(r,p)}$, $\ol{\alpha\beta}{(r,p)}$ are known as 
so-called $\mOmega$-integrals \cite{c29}. The $\mOmega$-integrals appear in 
calculation of collision integrals (integral parentheses \cite{c29}) in 
kinetic theory of gases, liquids and plasmas. The explicit form of 
$\mOmega$-integrals depends on type of interparticle interaction, that is 
taken into consideration. In our case, the total interparticle interaction 
potential consists on hard sphere and Coulomb parts. In such a way, two 
types of $\mOmega$-integrals will appear,
\bean
\lefteqn{\ds\ohs{\alpha\beta}{(r,p)}}\qquad\quad\;
&=&\intl_0^\infty\d y\;\e^{-y^2}y^{2p+3}\;
	\ohs{\alpha\beta}{(r)},\nonumber\\
\lefteqn{\ds\ol{\alpha\beta}{(r,p)}}\qquad\quad\;
&=&\intl_0^\infty\d y\;\e^{-y^2}y^{2p+3}\;
	\ol{\alpha\beta}{(r)},\nonumber\\
\lefteqn{\ds\ohs{\alpha\beta}{(r)}}\qquad\quad\;
&=&2\pi\intl_0^{\sigma_{\alpha\beta}}
	\d b\;b\ls 1-\cos^r\Big(\phantom{m}\!\!\!\!^{\alpha\beta}
	\chi'(b,y)\Big)\rs,\nonumber\\
\lefteqn{\ds\ol{\alpha\beta}{(r)}}\qquad\quad\;
&=&2\pi\intl_{\sigma_{\alpha\beta}}^\infty
	\d b\;b\ls 1-\cos^r\Big(\phantom{m}\!\!\!\!^{\alpha\beta}
	\chi^{*}(b,y)\Big)\rs.\nonumber
\eean
Here $\chi'$ is the scattering angle of hard spheres after collision, 
$\chi^*$ is the scattering angle of charged particles. The analytical 
calculation of $\mOmega$-integrals presented in \ref{ac} 

Except the thermal and mutual diffusion coefficients in the first 
approximation, the stress tensor $\tensor{P}'(\bmv{r}_1;t)$ and heat flow 
vector $\bmv{q}^*(\bmv{r}_1;t)$ one can calculate also and obtain other 
transport coefficients. The calculations give:
\[
\tensor{P}(\bmv{r}_1;t)=P(\bmv{r}_1;t)\tensor{I}-
	\kappa\big(\boldsymbol\nabla:\bmv{V}(\bmv{r}_1;t)\big)-
	2\eta\tensor{S}(\bmv{r}_1;t),
\]
\[
\bmv{q}(\bmv{r}_1;t)=-\lambda\boldsymbol\nabla T(\bmv{r}_1;t)+
	\sum_{\alpha=1}^M\omega_\alpha\bmv{d}^\alpha.
\]
Here $\kappa$ is the total bulk viscosity coefficient of mixture:
\be
\kappa=\frac 89\sum_{\alpha,\beta=1}^M\sigma_{\alpha\beta}^4\;
	g_2^{\alpha\beta}(\sigma_{\alpha\beta}|n,\beta)
	n_\alpha n_\beta\frac{m*}{m_\beta}
	\sqrt{2\pi m^*kT}=\sum_{\alpha,\beta=1}^M
	\kappa_{\alpha\beta},\label{e3.2.12}
\ee
$\eta$ is the total shear viscosity coefficient:
\bea
\lefteqn{\ds\eta=\frac 35\kappa+\frac 12\sum_{\alpha=1}^M n_\alpha kT
	\times{}}\label{e3.2.13}\\
&&\ls 1+\frac{2\pi}{15}\sum_{\beta=1}^M n_\beta
	\sigma_{\alpha\beta}^3\;
	g_2^{\alpha\beta}(\sigma_{\alpha\beta}|n,\beta)
	\ls 1+\frac{m_\alpha B^\beta_0}{m_\beta B^\alpha_0}\rs\rs
	B^\alpha_0,\nonumber
\eea
$\lambda$ is the total thermal conductivity coefficient:
\bea
\lefteqn{\ds\lambda=\frac 32\sum_{\alpha,\beta=1}^Mkm_\alpha m_\beta
	\ls\frac 12(m_\alpha+m_\beta)-\frac 18
	\frac{(m_\alpha-m_\beta)^2}{m_\alpha+m_\beta}\rs^{-3}
	\kappa_{\alpha\beta}+{}}\nonumber\\
&&\frac 54\sum_{\alpha=1}^Mn_\alpha k\sqrt{\frac{2kT}{m_\alpha}}
	\times{}\label{e3.2.14}\\
&&\ls 1-\frac{A^\alpha_0}{A^\alpha_1}+
	\frac{\pi}{5}\sum_{\beta=1}^M n_\beta\sigma_{\alpha\beta}^3\;
	g_2^{\alpha\beta}(\sigma_{\alpha\beta}|n,\beta)
	\ls 1+\frac{m_\alpha^{3/2}A^\beta_1}{m_\beta^{3/2}A^\alpha_1}
	\rs\rs A^\alpha_1,\nonumber
\eea
\[
\omega_\alpha=\frac 54nn_\alpha\sqrt{\frac{2k^3T^3}{m_\alpha}}
	E^\alpha_0.
\]

There is important to consider transport coefficients (\ref{e3.2.10}) -- 
(\ref{e3.2.14}) in a limiting transition, when the number of kinds $M$ goes 
to unity. In such a case one must obtain results of theory for an 
one-component system of charged hard spheres in a neutralizing background 
\cite{c2}. This is easy to see if one puts
\[
\begin{array}{ll}
m_a=m_b=m,&\qquad E^\alpha_0=A^\alpha_0=0,\\
n_a=n_b=n/2,&\qquad B^\alpha_0=B^\beta_0=B_0.\\
\end{array}
\]
By this means, there are no diffusion and thermal diffusion coefficients in 
an one-component system due to the absence of thermodynamic diffusion forces 
$\bmv{d}^\alpha$. But the equality $A^\alpha_0=0$ appears as the Fredgolm 
condition.

\section{Numerical calculations}
\setcounter{equation}{0}

In our paper numerical calculations have been carried out for transport 
coefficients $D^{\alpha\beta}$ \refp{e3.2.10}, $D_{\rm T}^\alpha$ 
\refp{e3.2.11}, $\kappa$ \refp{e3.2.12} and $\eta$ \refp{e3.2.13}. We 
have chosen mixtures of Ar-Kr, Ar-Xe, Kr-Xe as objects of our calculations. 
Neutral and charged mixtures of these gases are appropriate well for the 
model of neutral and charged hard spheres, which is considered here. This is 
so because other mixtures can have complex molecular structures. The 
transport coefficients have been calculated at several fixed dimensionless 
densities $\mDelta$ in a wide physically real for these mixtures ranges of 
temperature ($\mDelta=\sum_{\alpha=1}^M\mDelta_\alpha\sigma^3_\alpha$, 
$\mDelta_\alpha=\pi n_\alpha/6$, $\sigma$ is a hard sphere diameter).

Magnitudes for neutral systems ($Z_\alpha=0$) of some transport coefficients 
($D^{\alpha\beta}$, $\eta$) of Ar-Kr mixture have been compared with results 
of MD \cite{c30,c31} and HSED theory \cite{c32} for two fixed temperature 
points $T_1=508$ K and $T_2=1196$ K at the same dimensionless density 
$\mDelta=0.1$ ($n_\alpha=n_\beta$ of course). 
The coincidence of these values is exact (see Table \ref{t1}). 
\begin{table}[h]
\caption{The comparison of theoretical calculations of total shear 
viscosity $\eta$ and mutual diffusion $D^{\alpha\beta}$ with results of 
molecular dynamics simulations and other theories for Lenard-Jones liquids 
of argon and krypton. * -- data was borrowed from [32], ** -- data was 
borowed from [31], *** -- data was borrowed from [33].}
\vspace*{2ex}
\setlength{\tabcolsep}{1.75mm}
\begin{centering}
\footnotesize
\begin{tabular}{crlllll}\hline\hline\\
$\mDelta$&$T$, K&$\eta$, Pa$\cdot$s, MD&
	$\eta$, Pa$\cdot$s, {\footnotesize EDHST}&
	$\eta$, Pa$\cdot$s&$D^{\rm Ar-Kr}$, $\frac{\rm m^2}{\rm s}$, MD&
	$D^{\rm Ar-Kr}$, $\frac{\rm m^2}{\rm s}$\\\\\hline\\
0.1& 508&6{.}47$\cdot$10$^{-5\;*}$&&
	2{.}58$\cdot$10$^{-5}$&1{.}518$\cdot$10$^{-7\;*}$&
	2{.}568$\cdot$10$^{-7}$\\
0.1&1196&7{.}72$\cdot$10$^{-5\;**}$&8{.}56$\cdot$10$^{-5\;***}$&
	3{.}95$\cdot$10$^{-5}$&&\\\\\hline\hline
\end{tabular}\\
\end{centering}
\label{t1}
\end{table}

Values of shear viscosity of argon component $\eta_{\mbox{Ar}}$ of neutral 
mixture have been compared with results of \cite{c33,c34,c35,c36}. Theory 
parameters $\sigma_\alpha$ were taken from \cite{c37,c38,c39,c40,c41}. 
The obtained results are in a good agreement with 
\cite{c33,c34,c35,c36,c37,c38,c39,c40,c41} data. All obtained temperature 
dependencies of {\em all} transport coefficients at the absence and presence 
of long-range Coulomb interparticle interactions were published in our 
recent preprint \cite{preprint}. But in the present paper we give the 
combination of some numbers of them only, which reflect good a general 
tendency. The difference up to two orders in magnitudes for the values of 
the presented transport coefficients is explained as follows: firstly, 
different kind of interactions; secondly, different densities. Both these 
factors affect on magnitudes. This testifies that the obtained formulae are 
very sensitive on change of system parameters. 

Graphs of transport coefficients $D^{\alpha\beta}$, $D_{\rm T}^\alpha$ and 
$\eta$ in the case of their temperature dependencies are shown in Figure 1, 
kind concentrations are the same ($n_\alpha=n_\beta$). Each graph is 
assembled by pairs: neutral mixture ($Z_\alpha=0$) left and charged mixture 
($Z_\alpha\ne 0$) right. This allows to inspect changes in temperature 
behaviours of transport coefficients visually ($\sim T^{1/2}$ in Enskog-like 
theories and $\sim T^{3/2}$ at presence of long-range Coulomb interparticle 
interaction). The bulk viscosity $\kappa$ does not depend on charge 
$Z_\alpha$ and $T$. In such a way it does not change the nature of its 
temperature behaviours. 

A somewhat other quantitatively pattern appears when $n_\alpha\ne n_\beta$. 
If weighty component of mixture is in greater amount, partial lines in 
figures become much discrete and vice versa. Moreover, at some correlations 
of densities they can cross and the influence of light component becomes 
predominant. But temperature increasing diminishes this difference. (More 
figures see in details on www-server: 
www.icmp.lviv.ua/icmp/preprints/PS/9621Ups.gz).

At the end of this section one should add a little comment about choosing 
$g_2^{\alpha\beta}(r)$ and $D$, since they appear in transport coefficient 
expressions. As have been obtained earlier in this paper, transport 
coefficients take an analytical structure within the Chapman-Enskog method. 
But obtaining of any similar relations for $g_2^{\alpha\beta}(r)$ and $D$, 
which are perceived as being so adapted for considering system closely, is a 
completely different problem. Thus, in the present paper we have been used 
results done already. The quantity $D=1/2\mGamma$ is chosen as a solution 
to the next equation:
\[
4\mGamma^{2}=\frac{e^2}{\veps}\sum_{\alpha=1}^Mn_\alpha X_\alpha^2,
\]
\[
X_\alpha=\ls Z_\alpha-\frac{\pi}{2}\frac{\sigma_\alpha^2}{1-\mDelta}P_m\rs
	\ls 1+\mGamma\sigma_\alpha\rs^{-1},
\]
\[
P_m=\sum_{\alpha=1}^M\frac{n_\alpha\sigma_\alpha Z_\alpha}
	{1+\mGamma\sigma_\alpha}
	\ls 1+\frac{\pi}{2(1-\mDelta)}\sum_{\alpha=1}^M
	\frac{n_\alpha\sigma_\alpha^3}{1+\mGamma\sigma_\alpha}\rs^{-1},
\]
%
%
%
%
which has been proposed in \cite{c42}. But the expression for 
$g_2^{\alpha\beta}(r)$ has been assembled on the basis of results in 
\cite{c43,c44,c45,c46}. In \cite{c43}, the Percus-Yevick equation for the 
radial distribution function in liquid has been generalized to 
multicomponent mixtures, in \cite{c44} $g_2^{\alpha\beta}(r)$ has been 
calculated for ion mixtures at the presence of Coulomb and screened Coulomb 
interactions. A much precise formula for calculation of 
$g_2^{\alpha\beta}(r)$ on the basis of the Percus-Yevick approximation has 
been obtained in \cite{c45}, as well as a new closure for the 
Ornstein-Zernike equation for a system of charged particles with specific 
interparticle interaction potential has been used in \cite{c46}. Theoretical 
solutions there are compared with Monte Carlo simulations. Finally, we 
conclude that the following structure of $g_2^{\alpha\beta}(r)$ on hard 
sphere contact is the most plausible (in the present version of our work we 
wish to obtain analytical formulae only, choosing actual expressions for 
$g_2^{\alpha\beta}$:
\bean
g_2^{\alpha\alpha}(\sigma_\alpha)&=&
	\lc\lp 1+\frac{\mDelta}{2}\rp+
	\frac 32\mDelta_\beta\sigma_\beta^2(\sigma_\alpha-\sigma_\beta)
	\rc(1-\mDelta)^{-2},\\
g_2^{\beta\beta}(\sigma_\beta)&=&
	\lc\lp 1+\frac{\mDelta}{2}\rp+
	\frac 32\mDelta_\alpha\sigma_\alpha^2(\sigma_\beta-\sigma_\alpha)
	\rc(1-\mDelta)^{-2},\\
g_2^{\alpha\beta}(\sigma_{\alpha\beta})&=&
	\ls \sigma_\beta g_2^{\alpha\alpha}(\sigma_\alpha)+
	\sigma_\alpha g_2^{\beta\beta}(\sigma_\beta)\rs
	/2\sigma_{\alpha\beta},\\
g_2^{\beta\alpha}(\sigma_{\beta\alpha})&=&
	\ls \sigma_\beta g_2^{\alpha\alpha}(\sigma_\alpha)+
	\sigma_\alpha g_2^{\beta\beta}(\sigma_\beta)\rs
	/2\sigma_{\alpha\beta},\\
\sigma_{\alpha\beta}&=&\frac 12\lp\sigma_\alpha+\sigma_\beta\rp.
\eean
Structurally such a configuration is of simple design also.

\section*{Acknowledgements}

This work was supported partially by the State Fund for Fundamental 
Investigations at Ukrainian State Committee for Sciences and 
Technology, Project No 2.3/371 (1992-1994).

\begin{appendix}
\renewcommand{\thesection}{Appendix \Alph{section}.}
\newpage
\section{Calculation of 
$\bmv{A}_{\boldsymbol n}^{\boldsymbol\alpha}$-coefficients}\label{aa}
\setcounter{equation}{0}

The two mutually exchangeable infinite systems of equations for searching 
$A^\alpha_n$-co\-ef\-fi\-ci\-ents could be written in the form
\bean
\lefteqn{\ds\suml_{r=0}^{\infty}
	\ls\Big(\alpha_{sr}^{(a)}+\alpha_{sr}^{(ab,a)}\Big)A_r^a+
	\alpha_{sr}^{(ab,b)}A_r^b\rs=\beta_s^{(a)},}\\
\lefteqn{\ds\suml_{r=0}^{\infty}
	\ls\alpha_{sr}^{(ba,a)}A_r^a+
	\Big(\alpha_{sr}^{(b)}+\alpha_{sr}^{(ba,b)}\Big)A_r^b\rs=
	\beta_s^{(b)},}
\eean
where quantities $\alpha_{sr}$, $\beta_s$ have an integral structure and 
are expressed via $L^r_n$ polynomials. General expressions for these 
quantities are very cumbersome and their forms depend on the structure of 
collision integrals. But the general form is always the same and is 
completely shown in \cite{c27}. In our case
{\arraycolsep=0pt
\bean
\alpha_{sr}^{(a)}=\int\d\bmv{v}_1^a\d\bmv{v}_2^a{g}^{aa}\d\omega\;
	f_1^{a(0)}(\bmv{c}_1^a)f_1^{a(0)}(\bmv{c}_2^a)
	\lp\frac{m_a}{2kT}\rp^{3/2}
	L_s^{3/2}\lp\frac{m_a\lp{c}_1^a\rp^2}{2kT}\rp\times{}\\
\Bigg\{g_2^{aa}\Bigg[\lp\bmv{c}_1^a,\bmv{c}_1^{a\prime}\rp
	L_r^{3/2}\lp\frac{m_a\lp{c}_1^{a\prime}\rp^2}{2kT}\rp+
	\lp\bmv{c}_1^a,\bmv{c}_2^{a\prime}\rp
	L_r^{3/2}\lp\frac{m_a\lp{c}_2^{a\prime}\rp^2}{2kT}\rp-{}\\
\lp\bmv{c}_1^a,\bmv{c}_1^{a}\rp
	L_r^{3/2}\lp\frac{m_a\lp{c}_1^{a}\rp^2}{2kT}\rp-
	\lp\bmv{c}_1^a,\bmv{c}_2^{a}\rp
	L_r^{3/2}\lp\frac{m_a\lp{c}_2^{a}\rp^2}{2kT}\rp\Bigg]+{}\\
\lp\bmv{c}_1^a,\bmv{c}_1^{a*}\rp
	L_r^{3/2}\lp\frac{m_a\lp{c}_1^{a*}\rp^2}{2kT}\rp+
	\lp\bmv{c}_1^a,\bmv{c}_2^{a*}\rp
	L_r^{3/2}\lp\frac{m_a\lp{c}_2^{a*}\rp^2}{2kT}\rp-{}\\
\lp\bmv{c}_1^a,\bmv{c}_1^{a}\rp
	L_r^{3/2}\lp\frac{m_a\lp{c}_1^{a}\rp^2}{2kT}\rp-
	\lp\bmv{c}_1^a,\bmv{c}_2^{a}\rp
	L_r^{3/2}\lp\frac{m_a\lp{c}_2^{a}\rp^2}{2kT}\rp\Bigg\},
\eean
\bean
\alpha_{sr}^{(ab,a)}=\int\d\bmv{v}_1^a\d\bmv{v}_2^b{g}^{ab}\d\omega\;
	f_1^{a(0)}(\bmv{c}_1^a)f_1^{b(0)}(\bmv{c}_2^b)
	\lp\frac{m_a}{2kT}\rp^{3/2}
	L_s^{3/2}\lp\frac{m_a\lp{c}_1^a\rp^2}{2kT}\rp\times{}\\
\Bigg\{g_2^{ab}\Bigg[\lp\bmv{c}_1^a,\bmv{c}_1^{a\prime}\rp
	L_r^{3/2}\lp\frac{m_a\lp{c}_1^{a\prime}\rp^2}{2kT}\rp-
	\lp\bmv{c}_1^a,\bmv{c}_1^{a}\rp
	L_r^{3/2}\lp\frac{m_a\lp{c}_1^{a}\rp^2}{2kT}\rp\Bigg]+{}\\
\lp\bmv{c}_1^a,\bmv{c}_1^{a*}\rp
	L_r^{3/2}\lp\frac{m_a\lp{c}_1^{a*}\rp^2}{2kT}\rp-
	\lp\bmv{c}_1^a,\bmv{c}_1^{a}\rp
	L_r^{3/2}\lp\frac{m_a\lp{c}_1^{a}\rp^2}{2kT}\rp\Bigg\},
\eean
\bean
\alpha_{sr}^{(ab,b)}=\int\d\bmv{v}_1^a\d\bmv{v}_2^b{g}^{ab}\d\omega\;
	f_1^{a(0)}(\bmv{c}_1^a)f_1^{b(0)}(\bmv{c}_2^b)
	\frac{m_a\sqrt{m_b}}{(2kT)^{3/2}}
	L_s^{3/2}\lp\frac{m_a\lp{c}_1^a\rp^2}{2kT}\rp\times{}\\
\Bigg\{g_2^{ab}\Bigg[\lp\bmv{c}_1^a,\bmv{c}_2^{b\prime}\rp
	L_r^{3/2}\lp\frac{m_b\lp{c}_2^{b\prime}\rp^2}{2kT}\rp-
	\lp\bmv{c}_1^a,\bmv{c}_2^{b}\rp
	L_r^{3/2}\lp\frac{m_b\lp{c}_2^{b}\rp^2}{2kT}\rp\Bigg]+{}\\
\lp\bmv{c}_1^a,\bmv{c}_2^{b*}\rp
	L_r^{3/2}\lp\frac{m_b\lp{c}_2^{b*}\rp^2}{2kT}\rp-
	\lp\bmv{c}_1^a,\bmv{c}_2^{b}\rp
	L_r^{3/2}\lp\frac{m_b\lp{c}_2^{b}\rp^2}{2kT}\rp\Bigg\},
\eean
\footnotesize
\[
-\beta_s^{(a)}=\int\d\bmv{v}_1^af_1^{a(0)}(\bmv{c}_1^a)
	\frac{m_a\lp{c}_1^a\rp^2}{2kT}
	L_s^{3/2}\lp\frac{m_a\lp c_1^a\rp^2}{2kT}\rp\times{}
\]
$$
\Bigg\{\frac 52+
\left.\frac{m_a\!\!\lp{c}_1^a\rp^2}{2kT}
	\ls 1+\frac{2\pi}{5m_a}\sum_{\alpha=1}^Mm_\alpha n_\alpha
	\sigma_{a\alpha}g_2^{a\alpha}\rs\!\!+
	\frac \pi 3\sum_{\alpha=1}^Mn_\alpha\sigma_{a\alpha}^3g_2^{a\alpha}+
	\frac{2\pi m_a}{3\rho}\!\!\sum_{\alpha,\beta=1}^M
	n_\alpha n_\beta\sigma_{\alpha\beta}^3g_2^{\alpha\beta}\rc\!.
$$}

Let us consider the case, when indices $r$ and $s$ take two values only: 
zero and unity, which correspond to the first polynomial approximation. In 
such a case the system of equations becomes finite. It is convenient to 
represent it in the matrix form:
\[
\hat{a}_{ij}\hat{A}_j=\hat{b}_i,
\]
\[
\hat{A}_j=\mbox{col}\;\Big(A^a_0,\;A^a_1,\;A^b_0,\;A^b_1\Big),\quad
\hat{b}_i=\mbox{col}\;
\Big(\beta_0^{(a)},\;\beta_1^{(a)},\;\beta_0^{(b)},\;\beta_1^{(b)}\Big),
\]
matrix $\hat{a}_{ij}$ reads
\newcommand{\bc}{\!\!\!\!\!}
{\footnotesize
$$
\ls
\ba{llllllll}
\alpha_{00}^{(ab,a)}&\bc=a_{11}&\alpha_{01}^{(ab,a)}&\bc=a_{12}&
	\alpha_{00}^{(ab,b)}&\bc=a_{13}&\alpha_{01}^{(ab,b)}&\bc=a_{14}\\
\alpha_{10}^{(ab,a)}&\bc=a_{21}&\alpha_{11}^{(ab,a)}+\alpha_{11}^{(a)}
	&\bc=a_{22}&
	\alpha_{10}^{(ab,b)}&\bc=a_{23}&\alpha_{11}^{(ab,b)}&\bc=a_{24}\\
\alpha_{00}^{(ba,a)}&\bc=a_{31}&\alpha_{01}^{(ba,a)}&\bc=a_{32}&
	\alpha_{00}^{(ba,b)}&\bc=a_{33}&\alpha_{01}^{(ba,b)}&\bc=a_{34}\\
\alpha_{10}^{(ba,a)}&\bc=a_{41}&\alpha_{11}^{(ba,a)}&\bc=a_{42}&
	\alpha_{10}^{(ba,b)}&\bc=a_{43}&\alpha_{11}^{(ba,b)}+\alpha_{11}^{(b)}
	&\bc=a_{44}\\
\ea\rs.
$$}
The solution is written using the Kramers' method at once:
\renewcommand{\bc}{\!\!\!}
$$
A^a_0=\frac{1}{\det\hat{a}_{ij}}
\left|
\ba{llll}
b_1&\bc a_{12}&\bc a_{13}&\bc a_{14}\\
b_2&\bc a_{22}&\bc a_{23}&\bc a_{24}\\
b_3&\bc a_{32}&\bc a_{33}&\bc a_{34}\\
b_4&\bc a_{42}&\bc a_{43}&\bc a_{44}\\
\ea\right|,\;
A^a_1=\frac{1}{\det\hat{a}_{ij}}
\left|
\ba{llll}
a_{11}&\bc b_1&\bc a_{13}&\bc a_{14}\\
a_{21}&\bc b_2&\bc a_{23}&\bc a_{24}\\
a_{31}&\bc b_3&\bc a_{33}&\bc a_{34}\\
a_{41}&\bc b_4&\bc a_{43}&\bc a_{44}\\
\ea\right|,
$$
$$
A^b_0=\frac{1}{\det\hat{a}_{ij}}
\left|
\ba{llll}
a_{11}&\bc a_{12}&\bc b_1&\bc a_{14}\\
a_{21}&\bc a_{22}&\bc b_2&\bc a_{24}\\
a_{31}&\bc a_{32}&\bc b_3&\bc a_{34}\\
a_{41}&\bc a_{42}&\bc b_4&\bc a_{44}\\
\ea\right|,\;
A^b_1=\frac{1}{\det\hat{a}_{ij}}
\left|
\ba{llll}
a_{11}&\bc a_{11}&\bc a_{13}&\bc b_1\\
a_{21}&\bc a_{21}&\bc a_{23}&\bc b_2\\
a_{31}&\bc a_{31}&\bc a_{33}&\bc b_3\\
a_{41}&\bc a_{41}&\bc a_{43}&\bc b_4\\
\ea\right|.
$$
Quantities $a_{ij}$ can be calculated and represented via so-called 
$\mOmega$-integrals (see \ref{ac}). Then one obtains:
\[
a_{11}=-n_an_b\sqrt{\frac{32}{\pi}}\frac{m^*}{m_a}
	\lp g_2^{ab}(\sigma_{ab}|n,\beta)\ohs{aa}{(1,1)}+\ol{aa}{(1,1)}\rp,
\]
\bean
\lefteqn{\ds a_{12}=a_{21}=-n_an_b\sqrt{\frac{32}{\pi}}
	\lp\frac{m^*}{m_a}\rp^2\times{}}\\
\lefteqn{\ds\lp g_2^{ab}(\sigma_{ab}|n,\beta)
	\ls\frac 52\ohs{aa}{(1,1)}-\ohs{aa}{(1,2)}\rs+
	\frac 52\ol{aa}{(1,1)}-\ol{aa}{(1,2)}\rp,}
\eean
\[
a_{13}=4n_an_b\sqrt{\frac{m^*}{\pi m_b}}
	\lp g_2^{ab}(\sigma_{ab}|n,\beta)\ohs{ab}{(1,1)}+\ol{ab}{(1,1)}\rp,
\]
\bean
\lefteqn{\ds a_{14}=a_{23}=4n_an_b\sqrt{\pi}
	\lp\frac{m^*}{m_b}\rp^{3/2}\times{}}\\
\lefteqn{\ds\lp g_2^{ab}(\sigma_{ab}|n,\beta)
	\ls\frac 52\ohs{ab}{(1,1)}-\ohs{ab}{(1,2)}\rs+
	\frac 52\ol{ab}{(1,1)}-\ol{ab}{(1,2)}\rp,}
\eean
\bean
\lefteqn{\ds a_{22}=-n_an_b\sqrt{\frac{32}{\pi}}
	\lp\frac{m^*}{m_a}\rp\times{}}\\
\lefteqn{\ds\lp g_2^{ab}(\sigma_{ab}|n,\beta)
	\ls\lc\frac{30}{4}\lp\frac{m^*}{m_b}\rp^2+
	\frac{25}{4}\lp\frac{m^*}{m_a}\rp^2\rc
	\ohs{aa}{(1,1)}-{}\right.\right.}\\
\lefteqn{\ds\left.5\lp\frac{m^*}{m_a}\rp^2\ohs{aa}{(1,2)}+
	\lp\frac{m^*}{m_a}\rp^2\ohs{aa}{(1,3)}+
	\frac{2m_am_b}{(m_a+m_b)^2}\ohs{aa}{(2,2)}\rs+{}}\\
\lefteqn{\ds\ls\lc\frac{30}{4}\lp\frac{m^*}{m_b}\rp^2+
	\frac{25}{4}\lp\frac{m^*}{m_a}\rp^2\rc
	\ol{aa}{(1,1)}-{}\right.}\\
\lefteqn{\ds\left.\left.5\lp\frac{m^*}{m_a}\rp^2\ol{aa}{(1,2)}+
	\lp\frac{m^*}{m_a}\rp^2\ol{aa}{(1,3)}+
	\frac{2m_am_b}{(m_a+m_b)^2}\ol{aa}{(2,2)}\rs\rp+{}}\\
\lefteqn{\ds n_a^2\sqrt{\frac{8}{\pi}}\lp g_2^{aa}(\sigma_{ab}|n,\beta)
	\ohs{aa}{(2,2}+\ol{aa}{(2,2}\rp,}
\eean
\bean
\lefteqn{\ds a_{24}=4n_an_b\frac{1}{\sqrt{\pi}}
	\frac{(m^*)^{5/2}}{m_am_b^{3/2}}\times{}}\\
\lefteqn{\ds\lp g_2^{ab}(\sigma_{ab}|n,\beta)
	\lp\frac{55}{4}\ohs{ab}{(1,1)}-5\ohs{ab}{(1,2)}+
	\ohs{ab}{(1,3)}-2\ohs{ab}{(2,2)}\rp+{}\right.}\\
\lefteqn{\ds\left.\frac{55}{4}\ol{ab}{(1,1)}-5\ol{ab}{(1,2)}+
	\ol{ab}{(1,3)}-2\ol{ab}{(2,2)}\rp.}
\eean
The remaining quantities $a_{ij}$ are expressed via previous ones by simple 
exchange of indices $a\to b$ and vice versa, $b\to a$:
\bean
a_{31}=a_{13}&|& _{a\to b,\;b\to a},\\
a_{32}=a_{41}=a_{14}&|& _{a\to b,\;b\to a},\\
a_{33}=a_{11}&|& _{a\to b,\;b\to a},\\
a_{34}=a_{43}=a_{12}&|& _{a\to b,\;b\to a},\\
a_{42}=a_{24}&|& _{a\to b,\;b\to a},\\
a_{44}=a_{22}&|& _{a\to b,\;b\to a}.
\eean
The matrix elements $b_i$ in the first polynomial approximation are calculated 
exactly:
\bean
\lefteqn{\ds b_1=\frac{5}{4}n_a\lp 1+\frac{2\pi}{5m_a}\sum_{\alpha=1}^s
	m_\alpha n_\alpha \sigma_{a\alpha}^3
	g_2^{a\alpha}(\sigma_{a\alpha}|n,\beta)\rp-{}}\\
&&\frac{n_a}{2}\lp\frac 52+\frac{\pi}{3}\sum_{\alpha=1}^s
	n_\alpha\sigma_{a\alpha}^3
	g_2^{a\alpha}(\sigma_{a\alpha}|n,\beta)+{}\right.\\
&&\left.\frac{2\pi
	m_a}{3\rho}\sum_{\alpha,\beta=1}^sn_\alpha n_\beta\sigma_{\alpha\beta}^3
	g_2^{\alpha\beta}(\sigma_{\alpha\beta}|n,\beta)\rp,
\eean
\[
b_2=-\frac{15}{4}n_a\lp 1+\frac{2\pi}{5m_a}\sum_{\alpha=1}^s
	m_\alpha n_\alpha \sigma_{a\alpha}^3
	g_2^{a\alpha}(\sigma_{a\alpha}|n,\beta)\rp,
\]
\bean
b_{3}=b_{1}&|& _{a\to b,\;b\to a},\\
b_{4}=b_{2}&|& _{a\to b,\;b\to a},\\
\eean
\[
\mbox{where}\quad\rho=\sum_{\alpha=1}^s\rho_\alpha=\sum_{\alpha=1}^s
	m_\alpha n_\alpha.
\]

\newpage
\section{Calculation of 
$\bmv{B}_{\boldsymbol n}^{\boldsymbol\alpha}$-coefficients}\label{ab}
\setcounter{equation}{0}

The two mutually exchangeable infinite systems of equations for searching 
$B^\alpha_n$-co\-ef\-fi\-ci\-ents could be written in the form
\bea
\lefteqn{\ds\suml_{r=0}^{\infty}
	\ls\Big(\gamma_{sr}^{(a)}+\gamma_{sr}^{(ab,a)}\Big)B_r^a+
	\gamma_{sr}^{(ab,b)}B_r^b\rs=\zeta_s^{(a)},}\nonumber\\
\lefteqn{\ds\suml_{r=0}^{\infty}
	\ls\gamma_{sr}^{(ba,a)}B_r^a+
	\Big(\gamma_{sr}^{(b)}+\gamma_{sr}^{(ba,b)}\Big)B_r^b\rs=
	\zeta_s^{(b)},}\nonumber
\eea
where quantities $\gamma_{sr}$, $\zeta_s$ have an integral structure and 
are expressed via $L^r_n$ polynomials. General expressions for these 
quantities are very cumbersome and their forms depend on the structure of 
collision integrals. But general form is always the same and is completely 
shown in \cite{c27}. In our case
{\arraycolsep=0pt
\footnotesize
\[
\gamma_{sr}^{(a)}=\int\d\bmv{v}_1^a\d\bmv{v}_2^a{g}^{aa}\d\omega\;
	f_1^{a(0)}(\bmv{c}_1^a)f_1^{a(0)}(\bmv{c}_2^a)
	\lp\frac{m_a}{2kT}\rp^2
	L_s^{5/2}\lp\frac{m_a\lp{c}_1^a\rp^2}{2kT}\rp
	\bmv{c}_1^a\bmv{c}_1^a\times{}
\]
$$
\Bigg\{g_2^{aa}\Bigg[
	L_r^{5/2}\lp\frac{m_a\lp{c}_1^{a\prime}\rp^2}{2kT}\rp\!\!
	\lp\bmv{c}_1^{a\prime}\bmv{c}_1^{a\prime}-
	\frac 13\lp c_1^{a\prime}\rp^2\tensor{I}\rp+
	L_r^{5/2}\lp\frac{m_a\lp{c}_2^{a\prime}\rp^2}{2kT}\rp\!\!
	\lp\bmv{c}_2^{a\prime}\bmv{c}_2^{a\prime}-
	\frac 13\lp c_2^{a\prime}\rp^2\tensor{I}\rp-{}
$$
$$
L_r^{5/2}\lp\frac{m_a\lp{c}_1^a\rp^2}{2kT}\rp\!\!
	\lp\bmv{c}_1^{a}\bmv{c}_1^{a}-
	\frac 13\lp c_1^{a}\rp^2\tensor{I}\rp-
	L_r^{5/2}\lp\frac{m_a\lp{c}_2^a\rp^2}{2kT}\rp\!\!
	\lp\bmv{c}_2^{a}\bmv{c}_2^{a}-
	\frac 13\lp c_2^{a}\rp^2\tensor{I}\rp\Bigg]+{}
$$
$$
L_r^{5/2}\lp\frac{m_a\lp{c}_1^{a*}\rp^2}{2kT}\rp\!\!
	\lp\bmv{c}_1^{a*}\bmv{c}_1^{a*}-
	\frac 13\lp c_1^{a*}\rp^2\tensor{I}\rp+
	L_r^{5/2}\lp\frac{m_a\lp{c}_2^{a*}\rp^2}{2kT}\rp\!\!
	\lp\bmv{c}_2^{a*}\bmv{c}_2^{a*}-
	\frac 13\lp c_2^{a*}\rp^2\tensor{I}\rp-{}
$$
$$
L_r^{5/2}\lp\frac{m_a\lp{c}_1^a\rp^2}{2kT}\rp\!\!
	\lp\bmv{c}_1^{a}\bmv{c}_1^{a}-
	\frac 13\lp c_1^{a}\rp^2\tensor{I}\rp-
	L_r^{5/2}\lp\frac{m_a\lp{c}_2^a\rp^2}{2kT}\rp\!\!
	\lp\bmv{c}_2^{a}\bmv{c}_2^{a}-
	\frac 13\lp c_2^{a}\rp^2\tensor{I}\rp\Bigg\},
$$
\[
\gamma_{sr}^{(ab,a)}=\int\d\bmv{v}_1^a\d\bmv{v}_2^a{g}^{ab}\d\omega\;
	f_1^{a(0)}(\bmv{c}_1^a)f_1^{b(0)}(\bmv{c}_2^b)
	\lp\frac{m_a}{2kT}\rp^2
	L_s^{5/2}\lp\frac{m_a\lp{c}_1^a\rp^2}{2kT}\rp
	\bmv{c}_1^a\bmv{c}_1^a\times{}
\]
$$
\Bigg\{g_2^{ab}\Bigg[
	L_r^{5/2}\lp\frac{m_a\lp{c}_1^{a\prime}\rp^2}{2kT}\rp\!\!
	\lp\bmv{c}_1^{a\prime}\bmv{c}_1^{a\prime}-
	\frac 13\lp c_1^{a\prime}\rp^2\tensor{I}\rp-
	L_r^{5/2}\lp\frac{m_a\lp{c}_1^a\rp^2}{2kT}\rp\!\!
	\lp\bmv{c}_1^{a}\bmv{c}_1^{a}-
	\frac 13\lp c_1^{a}\rp^2\tensor{I}\rp\Bigg]+{}
$$
$$
L_r^{5/2}\lp\frac{m_a\lp{c}_1^{a*}\rp^2}{2kT}\rp\!\!
	\lp\bmv{c}_1^{a*}\bmv{c}_1^{a*}-
	\frac 13\lp c_1^{a*}\rp^2\tensor{I}\rp-
	L_r^{5/2}\lp\frac{m_a\lp{c}_1^a\rp^2}{2kT}\rp\!\!
	\lp\bmv{c}_1^{a}\bmv{c}_1^{a}-
	\frac 13\lp c_1^{a}\rp^2\tensor{I}\rp\Bigg\},
$$
\[
\gamma_{sr}^{(ab,b)}=\int\d\bmv{v}_1^a\d\bmv{v}_2^a{g}^{ab}\d\omega\;
	f_1^{a(0)}(\bmv{c}_1^a)f_1^{b(0)}(\bmv{c}_2^b)
	\frac{m_am_b}{(2kT)^2}
	L_s^{5/2}\lp\frac{m_a\lp{c}_1^a\rp^2}{2kT}\rp
	\bmv{c}_1^a\bmv{c}_1^a\times{}
\]
$$
\Bigg\{g_2^{ab}\Bigg[
	L_r^{5/2}\lp\frac{m_b\lp{c}_2^{b\prime}\rp^2}{2kT}\rp\!\!
	\lp\bmv{c}_2^{b\prime}\bmv{c}_2^{b\prime}-
	\frac 13\lp c_2^{b\prime}\rp^2\tensor{I}\rp-
	L_r^{5/2}\lp\frac{m_b\lp{c}_2^b\rp^2}{2kT}\rp\!\!
	\lp\bmv{c}_2^{b}\bmv{c}_2^{b}-
	\frac 13\lp c_2^{b}\rp^2\tensor{I}\rp\Bigg]+{}
$$
$$
L_r^{5/2}\lp\frac{m_b\lp{c}_2^{b*}\rp^2}{2kT}\rp\!\!
	\lp\bmv{c}_2^{b*}\bmv{c}_2^{b*}-
	\frac 13\lp c_2^{b*}\rp^2\tensor{I}\rp-
	L_r^{5/2}\lp\frac{m_b\lp{c}_2^b\rp^2}{2kT}\rp\!\!
	\lp\bmv{c}_2^{b}\bmv{c}_2^{b}-
	\frac 13\lp c_2^{b}\rp^2\tensor{I}\rp\Bigg\},
$$
\bean
\zeta_s^{(a)}=-\int\d\bmv{v}_1^a\;f_1^{a(0)}(\bmv{c}_1^a)
	L_s^{5/2}\lp\frac{m_a\lp{c}_1^a\rp^2}{2kT}\rp
	\bmv{c}_1^a\bmv{c}_1^a\times{}\\
\Bigg\{\frac{m_a\lp c_1^a\rp^2}{3kT}\lp1+\frac{2\pi}{3n}
	\sum_{\alpha,\beta=1}^Mn_\alpha n_\beta\sigma^3_{\alpha\beta}
	g_2^{\alpha\beta}-\frac{2\pi}{5m_a}\sum_{\alpha=1}^M
	m_\alpha n_\alpha \sigma^3_{a\alpha}g_2^{a\alpha}\rp\tensor{I}+{}\\
\frac{2\pi}{3}\sum_{\alpha=1}^M\!\ls 
	n_\alpha\sigma^3_{a\alpha}g_2^{a\alpha}-\!
	\sum_{\beta=1}^M\frac{n_\alpha n_\beta}{n}\sigma^3_{\alpha\beta}
	g_2^{\alpha\beta}\rs\tensor{I}-
	\frac{m_a}{kT}\bmv{c}_1^a\bmv{c}_1^a\ls 1+\frac{4\pi}{15m_a}
	\sum_{\alpha=1}^Mm_\alpha n_\alpha\sigma^3_{a\alpha}g_2^{a\alpha}
	\rs\!\!\Bigg\}.
\eean
}

Let us consider the case, when indices $r$ and $s$ 
take one zero value only, which corresponds to the zeroth polynomial 
approximation. In such a case the system of equations becomes finite. It is 
convenient to represent it in the matrix form:
\[
\hat{c}_{ij}\hat{B}_j=\hat{d}_i,
\]
\[
\hat{B}_{j}=\mbox{col}\;\Big(B^a_0,\;B^b_0\Big),\quad
\hat{d}_{i}=\mbox{col}\;\Big(\zeta_0^{(a)},\;\zeta_0^{(b)}\Big),
\]
matrix $\hat{c}_{ij}$ reads
\[
\lp
\ba{llllll}
\gamma_{00}^{(ab,a)}+\gamma_{00}^{(a)}&=&c_{11}\quad&
	\gamma_{00}^{(ab,b)}&=&c_{12}\\
\gamma_{00}^{(ba,a)}&=&c_{21}\quad&
	\gamma_{00}^{(ba,b)}+\gamma_{00}^{(b)}&=&c_{22}\\
\ea\rp.
\]
The solution is written using the Kramers' method at once:
\[
B^a_0=\frac{1}{\det\hat{c}_{ij}}\;\left|
\ba{ll}
d_1&c_{12}\\
d_2&c_{22}\\
\ea\right|,\quad
B^b_0=\frac{1}{\det\hat{c}_{ij}}\;\left|
\ba{ll}
c_{11}&d_1\\
c_{21}&d_2\\
\ea\right|.
\]
Quantities $c_{ij}$ can be calculated and represented via so-called 
$\mOmega$-integrals (see \ref{ac}). Then one obtains:
\bean
\lefteqn{\ds c_{11}=-n_an_b\frac{8}{3}\sqrt{\frac{kT}{\pi m_a}}
	\frac{(m^*)^2}{m_am_b}\times{}}\\
\eean
$$
\lp g_2^{ab}(\sigma_{ab}|n,\beta)
	\ls 5\ohs{aa}{(1,1)}+\frac{3m_b}{2m_a}\ohs{aa}{(2,2)}\rs+
	5\ol{aa}{(1,1)}+\frac{3m_b}{2m_a}\ol{aa}{(2,2)}\rp
$$
\bean
&&{}+4n_a^2\sqrt{\frac{kT}{\pi m_a}}\lp g_2^{aa}(\sigma_{ab}|n,\beta)
	\ohs{aa}{(2,2)}+\ol{aa}{(2,2)}\rp,
\eean
\bean
\lefteqn{\ds c_{12}=\frac{8n_an_b\sqrt{2kTm^*}}{3(m_a+m_b)}\times{}}\\
\lefteqn{\ds\lp g_2^{ab}(\sigma_{ab}|n,\beta)
	\ls 5\ohs{ab}{(1,1)}-\frac{3}{2}\ohs{ab}{(2,2)}\rs+
	5\ol{ab}{(1,1)}-\frac{3}{2}\ol{ab}{(2,2)}\rp.}
\eean
The remaining quantities $c_{ij}$ are expressed via previous ones by simple 
exchange of indices $a\to b$ and vice versa, $b\to a$:
\bean
c_{21}=c_{12}&|& _{a\to b,\;b\to a},\\
c_{22}=c_{11}&|& _{a\to b,\;b\to a}.
\eean
The matrix elements $d_i$ in the zeroth polynomial approximation are 
calculated exactly:
\bean
\lefteqn{\ds d_1=n_a\frac{5}{2}\lp 1+\frac{2\pi}{3n}
	\sum_{\alpha,\beta=1}^s
	n_\alpha n_\beta\sigma_{\alpha\beta}^3
	g_2^{\alpha\beta}(\sigma_{\alpha\beta}|n,\beta)-{}\right.}\\
&&\left.\frac{2\pi}{3m_a}\sum_{\alpha=1}^s
	m_\alpha n_\alpha\sigma_{a\alpha}^3
	g_2^{a\alpha}(\sigma_{a\alpha}|n,\beta)\rp-{}\\
&&\frac{\pi n_a}{3}\sum_{\alpha=1}^s\ls n_\alpha\sigma_{a\alpha}^3
	g_2^{a\alpha}(\sigma_{a\alpha}|n,\beta)-\!\sum_{\beta=1}^s
	\frac{n_\alpha n_\beta}{n}\sigma_{\alpha\beta}^3
	g_2^{\alpha\beta}(\sigma_{\alpha\beta}|n,\beta)\rs\\
&&{}+\frac{117\pi n_a}{56}\lp 1+
	\frac{4\pi}{15m_a}\sum_{\alpha=1}^s
	m_\alpha n_\alpha\sigma_{a\alpha}^3
	g_2^{a\alpha}(\sigma_{a\alpha}|n,\beta)\rp,
\eean
\bean
d_{2}=d_{1}&|& _{a\to b,\;b\to a}.
\eean

\newpage
\section{$\bmv\mOmega$-integrals}\label{ac}
\setcounter{equation}{0}

To calculate $\mOmega$-integrals analytically, one should find first the 
expressions for scattering angles $\chi'$ and $\chi^*$. In general case 
both these angles are complicated functions of many factors:
\[
\chi^{\prime,*}=\pi-2b\intl_{r_0}^{\infty}\d r\; r^2\lp 1-\frac{b}{r^2}-
	\frac{2\mPhi^{\rm hs,\,l}(r)}{m^*g^2}\rp^{-1/2},
\]
here the meaning of $r_0$ is searched from the extremum condition
\[
r_0^2-b^2-\frac{2r_0^2}{m^*g^2}\mPhi^{\rm hs,\,l}(r_0)=0.
\]
In view of geometrical reasoning, one can calculate $\cos\lp\chi'\rp$ 
exactly \cite{c29}~(Figure~\ref{c1}):
\[
\ds\cos\lp{}^{\alpha\beta}\chi'/2\rp=b/\sigma_{\alpha\beta}.
\]
\renewcommand{\thefigure}{\Alph{section}\arabic{figure}}
\begin{figure}[htbp]
\unitlength=1pt
\begin{centering}
\begin{picture}(133,109)
\put(0,0){\epsffile[239 345 372 454]{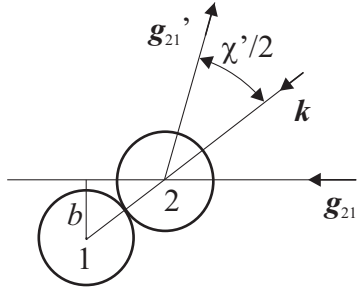}}
\end{picture}\\
\end{centering}
\caption{Collision of hard spheres.}
\label{c1}
\end{figure}
In the paper \cite{c25}, $\cos\lp\chi^*\rp$ is calculated approximately. In 
small scattering angles approximation one may be thought of as 
$|\bmv{g}'|\approx|\bmv{g}|$ and $\vartriangle\!\!\bmv{g}=\bmv{g}_y$. 
Considering the scattering dynamics as is shown in the drawing below (see 
Figure \ref{c2}), one may express $\cos\lp\chi^*\rp=\sin\theta$ 
($\chi^*+\Theta=\pi/2$) via $\vartriangle\!\!\bmv{g}$ and make expansion in 
the Teilor series. Here $\bmv{F}$ denotes the Coulomb force, which acts on 
particle 2 from the side of particle 1; in considered reference frame 
particle 1 is fixed while particle 2 impacts from infinity. Hence taking 
into account \refp{e3.7} one may obtain:
\[
\cos\lp{}^{\alpha\beta}\chi^*\rp=1-\frac 12
	\lp\frac{Z_\alpha Z_\beta e^2\pi}{2kTby^2}\rp^2.
\]
\begin{figure}[htbp]
\unitlength=1pt
\begin{centering}
\begin{picture}(353,200)
\put(0,51){\epsffile[197 321 415 470]{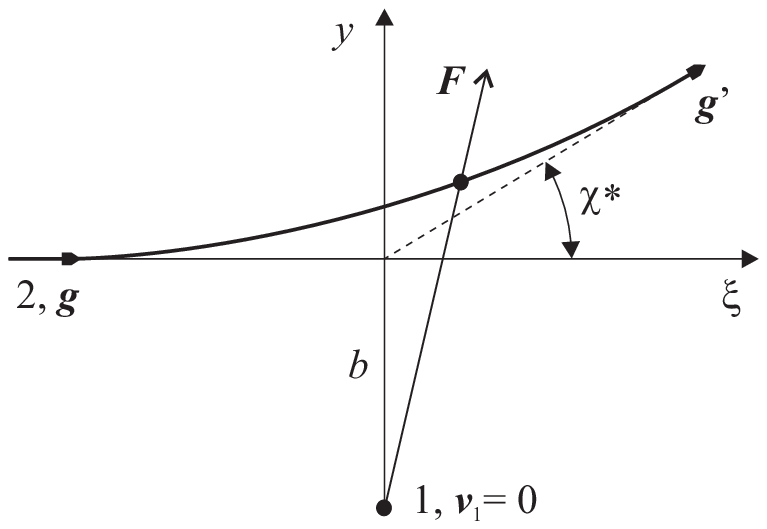}}
\put(190,0){\epsffile[224 341 387 450]{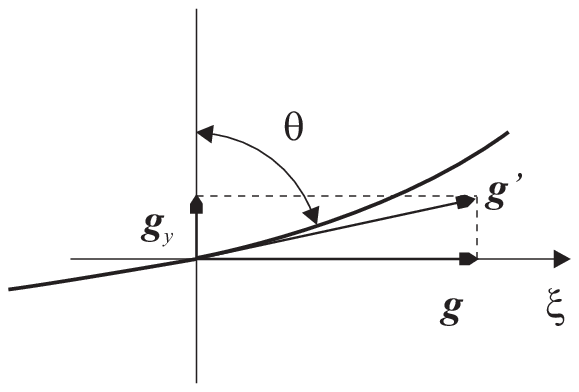}}
\end{picture}\\
\end{centering}
\caption{Scattering of two charged particles.}
\label{c2}
\end{figure}
This allows to write all $\mOmega$-integrals in an analytical form, which 
appear at searching $A^\alpha_n$-coefficients (see \ref{aa}), 
$B^\alpha_n$-coefficients (see \ref{ab}) and\\ $E^\alpha_n$-coefficients:
\[
\ba{ll}
\ds\ohs{\alpha\beta}{(1,1)}=\pi\sigma_{\alpha\beta}^2,&\qquad
\ds\ol{\alpha\beta}{(1,1)}=\frac{\pi^3}{2}
	\lp\frac{Z_\alpha Z_\beta e^2}{2kT}\rp^2
	\ln\frac{D}{\sigma_{\alpha\beta}},\\[1.5ex]
\ds\ohs{\alpha\beta}{(1,2)}=
	\phantom{1}3\ls\ohs{\alpha\beta}{(1,1)}\rs,&\qquad
\ds\ol{\alpha\beta}{(1,2)}=
	\phantom{1}2\ls\ol{\alpha\beta}{(1,1)}\rs,\\[1.5ex]
\ds\ohs{\alpha\beta}{(1,3)}=12\ls\ohs{\alpha\beta}{(1,1)}\rs,&\qquad
\ds\ol{\alpha\beta}{(1,3)}=12\ls\ol{\alpha\beta}{(1,1)}\rs,\\[1.5ex]
\ds\ohs{\alpha\beta}{(2,2)}=
	\phantom{1}2\ls\ohs{\alpha\beta}{(1,1)}\rs,&\qquad
\ds\ol{\alpha\beta}{(2,2)}=
	\phantom{1}4\ls\ol{\alpha\beta}{(1,1)}\rs,\\
\ea
\]
where the new quantity $D$ has been introduced, while calculating 
$\ol{\alpha\beta}{(r,p)}$. It is the cut-off radius of upper limit in the 
integral with respect to $b$. This is needed to avoid the difficulty of 
logarithmical divergency in $\mOmega_l$-integrals at long distances. The 
quantity $D$ has dimensionality of distance and meaning of screening radius 
(like a Debye radius). The equation for obtaining $D$ in any explicit form 
is a separate task, which will be considered in a next paper.

\end{appendix}

\vfill
\section*{Figure 1 caption}

\renewcommand{\Box}{\protect\rule{1mm}{1mm}}
\renewcommand{\circ}{\bullet}
\renewcommand{\triangle}{\mbox{\scriptsize$\blacktriangle$}}
\renewcommand{\times}{\mbox{\scriptsize$\blacklozenge$}}
The temperature dependencies of some transport coefficients for binary 
mixtures: a), c) and e) for neutral systems at $\mDelta=0.1$; b), d) and f) 
for once-ionized gases at $\mDelta=0.0125$. Everywhere $T$: [K]. a), b) are 
mutual diffusion coefficients for mixtures of Ar-Kr ($\circ$), Ar-Xe 
($\Box$), Kr-Xe ($\triangle$), $D^{\alpha\beta}$ a): 10$^{-7}$, b): 
10$^{-9}$[m$^2$/s]. c), d) are thermal diffusion coefficients for Ar-Kr 
mixture: argon's Ar ($+$) and krypton's Kr ($\times$) components, 
$D_{\rm T}$ c): 10$^{-5}$, d): 10$^{-8}$[kg/m$\cdot$s]. e), f) are shear 
viscosity coefficients for Ar-Kr mixture ($\circ$), argon's Ar ($\Box$) and 
krypton's Kr ($\triangle$) components, $\eta$ e): 10$^{-4}$, f): 
10$^{-6}$[Pa$\cdot$s].

\end{document}